\documentclass[10pt]{article}

\usepackage{cite}
\usepackage{url}
\usepackage{multicol} 
\usepackage[utf8]{inputenc}
\usepackage{graphicx}
\usepackage{float}

\begin{document}

\title{A novel approach to simulate gene-environment interactions in complex diseases}

\author{Roberto Amato$^{1,2\,\,\dag}$\footnote{Corresponding author: roamato@na.infn.it}%
      \and
         Michele Pinelli$^{1,3\,\,\dag}$
      \and
         Daniel D'Andrea$^1$
      \and
         Gennaro Miele$^{1,2,4}$
      \and
         Mario Nicodemi$^{1,4,5}$
      \and
         Giancarlo Raiconi$^{1,6}$
      \and
         Sergio Cocozza$^{1,3}$
      }

\date{}
\maketitle

\begin{scriptsize}\noindent(1) Gruppo Interdipartimentale di Bioinformatica e Biologia Computazionale, Universit\`a di Napoli ``Federico II'' - Universit\`a di Salerno, Italy\\
(2) Dipartimento di Scienze Fisiche, Universit\`a di Napoli ``Federico II'', Complesso Universitario di Monte S.Angelo, via Cinthia 6, 80126 - Napoli, Italy\\
(3) Dipartimento di Biologia e Patologia Cellulare e Molecolare ``L. Califano'', Universit\`a di Napoli ``Federico II'', via Pansini 5, 80131 - Napoli, Italy\\
(4) INFN Sezione di Napoli, Complesso Universitario di Monte S.Angelo, via Cinthia 6, 80126 - Napoli, Italy\\
(5) Complexity Science Center and Department of Physics, University of Warwick, UK\\
(6) Dipartimento di Matematica e Informatica, Universit\`a di Salerno, via ponte don Melillo, 84084 - Fisciano (SA), Italy\\
(\dag) These authors contributed equally to this work\end{scriptsize}

\begin{abstract}
\begin{description}
\textbf{Background:} Complex diseases are multifactorial traits caused by both genetic and environmental factors. They represent the most part of human diseases and include those with largest prevalence and mortality (cancer, heart disease, obesity, etc.). Despite of a large amount of information that have been collected about both genetic and environmental risk factors, there are relatively few examples of studies on their interactions in epidemiological literature. One reason can be the incomplete knowledge of the power of statistical methods designed to search for risk factors and their interactions in this data sets. An improving in this direction would lead to a better understanding and description of gene-environment interaction. To this aim, a possible strategy is to challenge the different statistical methods against data sets where the underlying phenomenon is completely known and fully controllable, like for example simulated ones.

\textbf{Results:} We present a mathematical approach that models gene-environment interactions. By this method it is possible to generate simulated populations having gene-environment interactions of any form, involving any number of genetic and environmental factors and also allowing non-linear interactions as epistasis. In particular, we implemented a simple version of this model in a Gene-Environment iNteraction Simulator (GENS), a tool designed to simulate case-control data sets where a one gene-one environment interaction influences the disease risk. The main effort has been to allow user to describe characteristics of population by using standard epidemiological measures and to implement constraints to make the simulator behavior biologically meaningful.

\textbf{Conclusions:} By the multi-logistic model implemented in GENS it is possible to simulate case-control samples of complex disease where gene-environment interactions influence the disease risk. The user has a full control of the main characteristics of the simulated population and a Monte Carlo process allows random variability. A Knowledge-based approach reduces the complexity of the mathematical model by using reasonable biological constraints and makes the simulation more understandable in biological terms. Simulated data sets can be used for the assessment of novel statistical methods or for the evaluation of the statistical power when designing a study.

\end{description}

\end{abstract}

\section{Background}

Complex Diseases (CD) are caused by variations in multiple loci interacting with each others and with environmental
factors \cite{Weeks1995}. Many complex traits, such as cancer, heart disease, obesity, diabetes, and many common psychiatric and neurological conditions, have large prevalence and mortality among
human diseases \cite{GAIN2007,Lohmueller2003}.

The concept of Gene-Environment interaction (GxE) is theoretically
central in CD \cite{Haldane1946}. It is widely accepted that GxE must be considered in CD to avoid a serious underestimations
of the disease risk and inconsistencies of replication among different studies. Furthermore, the action of consider the
GxE could focus medical intervention by identifying sub-groups of
individuals who are more sensible to specific environmental
exposures \cite{Hunter2005}. However there are very few examples
of well described GxE in scientific literature \cite{Khoury2005}.
Instead, a large amount of information have been collected about
both single genetic and environmental risk factors individually
taken, because the majority of the studies examined the main
effect of single factors instead of examining the interactions
\cite{Khoury2005,Hoh2003,WellcomeTrust2007}.

In our opinion, one reason of such a failure concerns the
statistical approach. Several statistical methods aimed to the
identification of factors' interactions have been described and
have been used to identify GxE, such as Logistic Regression \cite{Yosuke2004} and
MDR \cite{Ritchie2001,Hahn2003}. However, the
performances of these methods can be influenced by many variables
such as the sample size, the number of involved factors, the type
of interaction, the model of inheritance, the allelic frequencies,
the distributions of the environmental factors, and the relative
strength of the different factors on the disease. Unfortunately,
only for few real populations some of these characteristics are
known, and, however there are not enough real data sets to assess
the performances of statistical methods.

In this scenario, as an alternative approach one can imagine to
use simulated populations in order to assay the statistical power
of different methods. In population
genetics, there are several genetic data simulators (for a
complete list see \cite{geneticwebsite}) and some of them can also
simulate GxE. However, many of these tools are tailored to
simulate population evolution (as coalescent \cite{Gasbarra2005}
and forward-time methods \cite{Peng2007b}) or pedigrees  for
linkage analysis \cite{Leal2005,Schmidt2005}. Case-control
samples for complex diseases can be extracted by these data, for
example, by picking individuals in pedigree-based datasets (SIMLA
\cite{Schmidt2005}). However, such processes are not
straightforward, because these packages are not specifically
designed to this aim. In addition, by this approach the user has
not complete control of the parameters of the system.

To this aim, we propose a novel method, the Multi-Logistic Model,
that mathematically describes gene-environment interactions
similar to those in case-control studies. By this method is
possible to model GxE of any form, involving any number of genetic
and environmental factors, also allowing gene-gene interactions,
as epistasis. A simple version have been implemented in the
Gene-Environment iNteraction Simulator (GENS), a tool designed to
simulate case-control data sets where a one gene-one environment
interaction influences the disease risk. Moreover, we employed as
input common epidemiological measures to make easier the
simulation of data close to those from previous studies or
literature. This makes the tool more friendly to biomedical
community and allows to use as input data those extracted from
previous studies in literature.

\section{Results}

\subsection{The Multi-Logistic Model for gene-environment interaction}

The mathematical approach behind the simulation of the disease
risk involving GxE is based on a system of logistic relationships.
We called this approach Multi-Logistic Model (MLM) and
specifically designed it to describe disease risk in data set that
simulate case-control sample. In the simulated data set, each
individual has $G$ genetic factors and is exposed to $E$
environmental factors. Genetic factors are denoted by $g^a_{i_a}$
where $a=1,...,G$. The genetic factors are biallelic Single
Nucleotide Polymorphisms which result in three diploid genotypes,
namely the first homozygote (AA, $i_a=1$), the heterozygote (Aa,
$i_a=2$) and the second homozygote (aa, $i_a=3$). Genetic
frequencies for each factor are denoted by $P^G(g^a_{i_a})$ where
$\forall a$ $\sum_{i_a=1,2,3}P^G(g^a_{i_a})=1$. The environmental
variables, instead, are denoted by $x^b_{j_b}$, where $b=1,...,E$
and $j_b$ is an index which runs over the possible discretized
values of the variable $b$. They are characterized by
exposure probabilities denoted by $P^E(x^b_{j_b})$ (where again
$\forall b$ $\sum_{j_b}P^E(x^b_{j_b})=1$).

Let us consider a particular individual characterized by $(E+G)$ values of $x^b_{j_b}$ and $g^a_{i_a}$. In
general the disease risk $R$ is a function
of all of them. The disease risk for such an individual
$(g^a_{i_a}, x^b_{j_b})$ is defined by the conditioned probability
\begin{equation}
R(g^a_{i_a},x^b_{j_b}) = P\left(\mbox{affected}\left|g^a_{i_a},x^b_{j_b}\right)\right. \label{eq:R1}
\end{equation}
where $P\left(\mbox{affected}\left|g^a_{i_a},x^b_{j_b}\right)\right.$ is the probability of the individual to be \textit{affected}. In our model we assume a logistic expression for $R$:
\begin{equation}
R(g^a_{i_a},x^b_{j_b}) = \left[1+\exp\left\{\alpha_{i_1,..,i_G} + \sum_{b=1}^E\beta_{i_1,..,i_G}^b x^b_{j_b}\right\}\right]^{-1}\label{eq:R2}
\end{equation}
where $\alpha_{i_1,..,i_G}$ and $\beta_{i_1,..,i_G}^b$ are free parameters determined by the genetic factors and governing the shape of the function (Figure 1).

\subsection{Gene-Environment iNteraction Simulator}

We implemented the MLM in the Gene-Environment iNteraction
Simulator (GENS). For sake of simplicity we describe, in this
phase, a simple interaction between one genetic and one
environmental factor even though we continue to describe an
individual by assigning to him a $(E+G)$-tuple of
characteristics.

As consequence of this choice, the MLM gets a simpler form. In
particular, we can drop the indexes $a$ and $b$ in the expression
of disease risk (\ref{eq:R2}). Thus, by denoting with $g_i$ the
genotype of the chosen gene and with $x_j$ the exposure level of
the environmental factor involved, we have
\begin{equation}
R(g^a_{i_a},x^b_{j_b})= R(g_{i},x_j)= \left[1+\exp\left\{ \alpha_i + \beta_i \, x_j\right\}\right]^{-1}. \label{eq:1}
\end{equation}
In other words, the MLM reduces to three logistic functions, one
for each genotype.

It is possible to think of $\alpha_i$ as the basal genetic disease
risk in individuals with that genotype. The greater is $\alpha_i$
the stronger is the disease risk, independently of the
contribution of the environmental factor. In particular, for
vanishing $\alpha_i$ there is no basal risk and the risk is
totally ascribed to the environmental exposure ($x_j$).
Analogously, $\beta_i$ represents the coefficient associated to
the environmental exposure, thus the greater is $\beta_i$ the
greater risk is associated to an increasing of the environmental
exposure. In other words, $\beta_i$ models, for genotype $i$, the
sensibility to the environmental factor exposure. Consequently,
for vanishing $\beta_i$ the environmental exposure has no effect
on the disease risk.

To describe populations by standard epidemiological measures, we
implied the relative risk as the measure of the role of a genetic
factor on the disease risk. In particular, by defining the Total Risk
($TR$) in a specific genotype $i$ as
\begin{equation}
TR_i = \sum_{j} P^E(x_{j}) \left[1+\exp\left\{ \alpha_i + \beta_i \, x_j\right\}\right]^{-1}
\label{eq:5}
\end{equation}
(which holds under the hypothesis of independence among different
environmental variables) one can define the Relative Risk $RR_{kl} \equiv TR_k/TR_l$.

We take one homozygote as a reference point (say AA, denoted with
$i=1$), the other homozygote (say aa, $i=3$) has an equal or larger risk
than the first one, and the heterozygote (Aa, $i=2$) has a risk
ranging within the two homozygotes. Furthermore, we assume the
relative risk of heterozygote to be within those of the two
homozygotes ($1 \leq RR_{21} \leq RR_{31}$). In particular, if the
heterozygote risk is the same of the first homozygote it simulates
a recessive effect, else if the heterozygote has the same risk of
the second homozygote it simulates a dominant effect. Other
situations are called co-dominant.

Formally, the relative risk of heterozygote $RR_{21}$ is defined as
\begin{equation}
RR_{21} = \left(RR_{31}\right)^W
\end{equation}
where the $W$ lets to model various inheritance effects: recessive
($W=0$), dominant ($W=1$), and co-dominant ($0 < RR_{21} < 1$)
\cite{Schmidt2005}.

Marginal risk of the environmental factor is input as the odds ratio of the increase of one unit in the level of exposition. This value is then transformed in the coefficients $\beta_i$ of the multi-logistic model. Anyway at maximum only one $\beta_i$ is provided by the user, leaving the tool deriving other values to respect all the constraints.

\subsubsection{Type of GxE interaction}

To describe the GxE in biological understandable terms, we consider a
genetic and an environmental alone model and two models of
interactions involving both genetics and environment (Table 1 and
Figure 2). The first two models could be useful as reference.

In the first model, the \emph{Genetic Model} (GM), each subject
carrying a genotype has the same disease risk regardless of the
environmental exposure. This situation is modeled by giving
vanishing effect to the environmental variable, namely fixing all
the $\beta_i$ equal to zero. In the second model, the
\emph{Environmental Model} (EM), the risk is due to the
environmental exposure only. This situation is modeled by imposing
$\alpha_i$ and $\beta_i$ equal across the genotypes with a non-vanishing $\beta_i$. This choice provides the same risk
independently of the carried genotype.

The third model simulates the scenario where the \emph{gene
modules response to environment} (Gene Environment interaction
Model - GEM). In this case the genetics does not affect directly
the disease risk, but modules the response to the environmental
exposure. In other words, some genotypes are more prone than
others to develop the disease if exposed to the same environmental
level. For this interaction all the $\alpha_i$ are equal to each
others (no direct genetic effect) while $\beta_i$ are different.
The last effect is \emph{Additive Model} (AM), where genetic and
environment module directly, independently and additively the
risk, and the environmental exposure has the same effect in all
the genotypes (equal $\beta_i$). For this model, there are no
complex interactions between the genetics and the environmental
exposure, but the risk is in such a way the sum of the risk due to
genetics and environment. Of course the user can create further
type of GxE by freely imposing $\alpha_i$ and $\beta_i$.

\subsubsection{Knowledge-Aided Parametrization System}

To translate the population parameters into
coefficients of the MLM, we implemented the so-called
Knowledge-Aided Parametrization System (KAPS). This module
derives the values of $\alpha_i$ and $\beta_i$ starting from
genotype frequencies, relative risk and model of inheritance of
the genetic factor, distribution and odds ratio of the
environmental factor, type of GxE and proportion of affected
individual in the sample ($m$).

The key issue is that the overall disease frequency in the population $m$ is given by
\begin{equation}
\sum_i P^G(g_i) \left( \sum_j P^E(x_j) \left[ 1+\exp\left\{ \alpha_i + \beta_i \, x_j \right\} \right] ^{-1} \right) = m
\label{eq:4}
\end{equation}

Dividing Eq. \ref{eq:4} by $TR_1$ and by means of some algebraic manipulation, it is straightforward to show that
\begin{equation}
\sum_{j} \frac{P^E(x_{j})}{1+\exp\left\{ \alpha_1 + \beta_1
\, x_j\right\}} = \frac{m}{P^G(g_1)+P^G(g_2)RR_{21}+P^G(g_3)RR_{31}}
\label{eq:6}
\end{equation}
In a similar way it is possible to derive the expressions for the
marginal risk of the other genotypes. By numerically solving this
set of three equations (one for each $TR_i$) it is possible to
obtain $\alpha_i$ and $\beta_i$ coefficients that match at the
best the user's requests.

\subsubsection{Algorithm and Implementation}

The simulation procedure is divided into several steps (Figure 3). First of all,
the genotypes of $G$ genetic factors and the levels of exposure of
$E$ environmental factors are assigned to the $N$ individuals.
Consequently, the user of the simulator inputs the sample population
characteristics (Table 2) and hence the coefficients of the MLM
are computed. Finally, the disease risk and the disease status are
assigned to each individual.

Concerning the genetic factors, the user can provide the allelic
frequencies or let the simulator randomly select them (with an
uniform distribution between $0.1$ and $0.9$). In both cases, the
Hardy-Weinberg's law is used for the computation of the
frequencies of the genotypes. Afterward, by means of a Monte Carlo
method, the genotype of each genetic factor is randomly assigned
to each individual according to the genotypic frequencies.
Similarly for the environmental factors, the user can use a
distribution function, among a set of predefined ones, or provide
an empirical distribution $P^E(x^b_{j_b})$. Again by a Monte Carlo
process the exposures of environmental factors are assigned
according to distribution functions.

After the assignment of the genetic and environmental factors to
individuals, the next step is the assignment of the phenotype. To
this aim the system computes the coefficients of the MLM in order
to create the relationship between population characteristics,
type of GxE interaction, and disease risk. The actual computation
of the coefficients is performed by KAPS that, by means of Eq.
\ref{eq:6} and similar ones for $TR_2$ and $TR_3$, solves
numerically the resulting system of three equations and returns
$\alpha_1$,$\alpha_2$,$\alpha_3$,$\beta_1$,$\beta_2$ and
$\beta_3$.

The disease risk ($0 \leq R(g_{i},x_j) \leq 1$) is assigned by the
MLM (Eq. \ref{eq:1}) by using the parameters previously
identified. In particular, for each individual his genotype $i$
imposes the coefficients $\alpha_i$ and $\beta_i$ computed by
KAPS, while the exposure level is the value of the covariate
$x_j$. Last step is to assign a disease status (affected/not
affected) to the individuals. To this aim, again by a Monte Carlo
process, the system generates a random number with uniform
distribution in $[0, 1]$ and assigns to the individual the status
1 (affected) if this number is less then his risk $R(g_{i},x_j)$,
or 0 (not affected) otherwise.

An implementation of GENS is provided as a set of Matlab 7.0
scripts, which can be freely modified to address different
requirements (different risk function, multiloci interaction,
etc.).

\section{Discussion and Conclusions}

In this article we present a novel mathematical approach to model
GxE in complex diseases. This approach is based on a MLM and it is
specifically tailored to model disease risk in data set that
simulates case-control samples. We implemented this method in
Gene-Environment iNteraction Simulator (GENS), a tool designed to
yield case-control samples for GxE. These tools could be useful to
generate simulated data sets in order to assess the performances
of statistical methods.

The necessity to provide simulated populations is due to the
difficulty to obtain real populations in which enough parameters
are known to be related to the phenotype. Furthermore, during the
design of a statistical study, simulated populations can be also
used to estimate the expected statistical power when assuming
different types of GxE\cite{GarciaClosas1999}. To this aim, we
focused on the possibility to use as input characteristics
extracted by real populations (such as allelic frequencies,
environmental factors distributions, risk given by genetic and
environmental factors, etc.). In this way it is straightforward
also to make replica of real populations and evaluate the change
of statistical power due to changes of parameters as the
sample size ($N$), the disease frequency ($m$), the type of GxE,
etc.

The key idea underlying the MLM is the modeling of the disease
risk in each combination of genetic factors (genotypes) as a
different mathematical function of the environmental exposure
(Figure 1). In this way it is possible to model any type of
interaction between genetic and environmental factors, also
complex and non-linear ones. We based our approach on the logistic
function. This function is widely used in epidemiological studies
and has several advantages: it follows the Weber-Fechner law; 
as a risk factor naturally ranges from 0 to 1; well describes a saturation behavior for large values of
exposure \cite{Hosmer2000}. Moreover, the coefficients of the covariates correspond
to the logarithm of the odds ratio due to a one-unit increase (in
this case the environmental factor) \cite{Hosmer2000}. In
particular, to calculate disease risk genetic factors of
individuals set coefficients of the function while the
environmental factors assign a value to its covariates.

We implemented the MLM in the Gene-Environment iNteraction
Simulator (GENS), a GxE simulator for case-control studies. The
intended audience of GENS is the biomedical community, thus the
main efforts have been to describe populations by standard
epidemiological measures, to implement constraints to make the
simulator behavior biologically meaningful, and to define the GxE
in biological understandable terms.

In theory, the MLM can model many genetic-many environmental
factors interactions. However, for sake of simplicity we decided
to focus, in this phase, on an interaction between one genetic and
one environmental factor. In this way, by using only one gene-one
environmental factor interactions it is much easier to use as
input standard epidemiological measures. Nevertheless, even in this
simple situation, the handling of the interaction is not trivial.
Furthermore, in simulated populations beside the involved factors
there are other ones that are left as noisy background, as
frequently occurs in real data sets.

Even in this simpler scenario, modeling the desired
characteristics of a population can be very difficult except for
some particular and trivial cases mainly because it is necessary
to provide several coefficients to the mathematical model.
However, having several coefficients with a difficult
interpretation is a common pitfall when modeling complex
interactions. Therefore, to overcome this limitation we have
implemented the Knowledge-Aided Parametrization Subsystem (KAPS).
This system exploits a set of reasonable biological constraints to
reduce the complexity of the system. First of all, concerning the
genetic factors we imposed that the risk assigned to the
heterozygote falls between the two homozygotes. Secondly, we
adopted a \textit{qualitative} description of the GxE. In
particular, each type of GxE can be modeled as a set of equality
and inequality of $\alpha_i$ and $\beta_i$ among genotypes. We
pre-determined two types of GxE, an additive (AM) and a modulative
type (GEM). The user has only to select which type of GxE must be
simulated, without providing additional information. In this way,
we can reduce the complexity of the system and, thus, reduce the
grades of freedom of the mathematical model. Finally, KAPS solves
the system of equations to derive coefficients in order to
complain with both biological constraints and population
characteristics imposed by the user. As a consequence, to simulate
a population only classical epidemiological parameters have to be
provided (Table 2). However, the user can simulate interaction of
any kind since we left the possibility to input all the
coefficient of the MLM, and even to substitute the logistic
expression with a different one.

In population genetics, data simulation has been mainly used to
study population evolution and linkage disequilibrium, especially
in pedigree of Mendelian disease
\cite{Gasbarra2005,Peng2007b,Leal2005}. Although some tools have
been specifically designed to simulate data set of complex
diseases \cite{Schmidt2005}, they usually do not directly produce
case-control data sets. Indeed, these tools usually provide
family- or population-based data distributed across multiple
generations. On the contrary, GENS is specifically designed to
produce case-control data sets and to produce data set as closer
as possible to real cases in a simple manner. In addition,
differently from a {\it naive} logistic model, the MLM allows to
model non-liner phenomena such as epistasis.

One of the shortcoming of GENS compared to other tools is the
limit to one gene-one environment interaction even if this choice
have been made because would be easier to describe and understand
the joined and single role of the factors. Anyway, it should be
noted that this limit accounts mainly to the present
implementation, in particular to KAPS. In fact, the
multi-logistic model can be easily used to simulate many
genes-many environments interactions by applying Eq. \ref{eq:R2}
and providing enough coefficients. The number of environmental
factors is increased by adding additional covariates in the
functions to consider their effects. Instead, the number of
genetic factors involved in the disease risk is increased by
defining additional logistic functions in the multi-logistic
model. Furthermore, the multi-logistic model can be extended to
use different functions for each combination of genetic factors.

As our approach is widely based on a Monte Carlo process, the
system naturally takes into account the stochasticity present in
any real data set obeying to probabilistic laws, namely data sets
created with the same characteristics result to be randomly
different.

In conclusion, by the multi-logistic model and GENS is possible
to simulate case-control samples of complex diseases where
gene-environment interactions influence the disease risk. The user
has a full control of the main characteristics of the simulated
population and a Monte Carlo process allows random variability. A
knowledge-based approach reduces the complexity of the mathematical
model by using reasonable biological constraints and makes the
simulation more understandable in biological terms. Simulated data
sets can be used for the assessment of novel statistical methods
or for the evaluation of statistical power when designing a study.

\bibliographystyle{plain}
\bibliography{paper}

\section{Figures}
\begin{figure}[H]
\includegraphics[width=\textwidth]{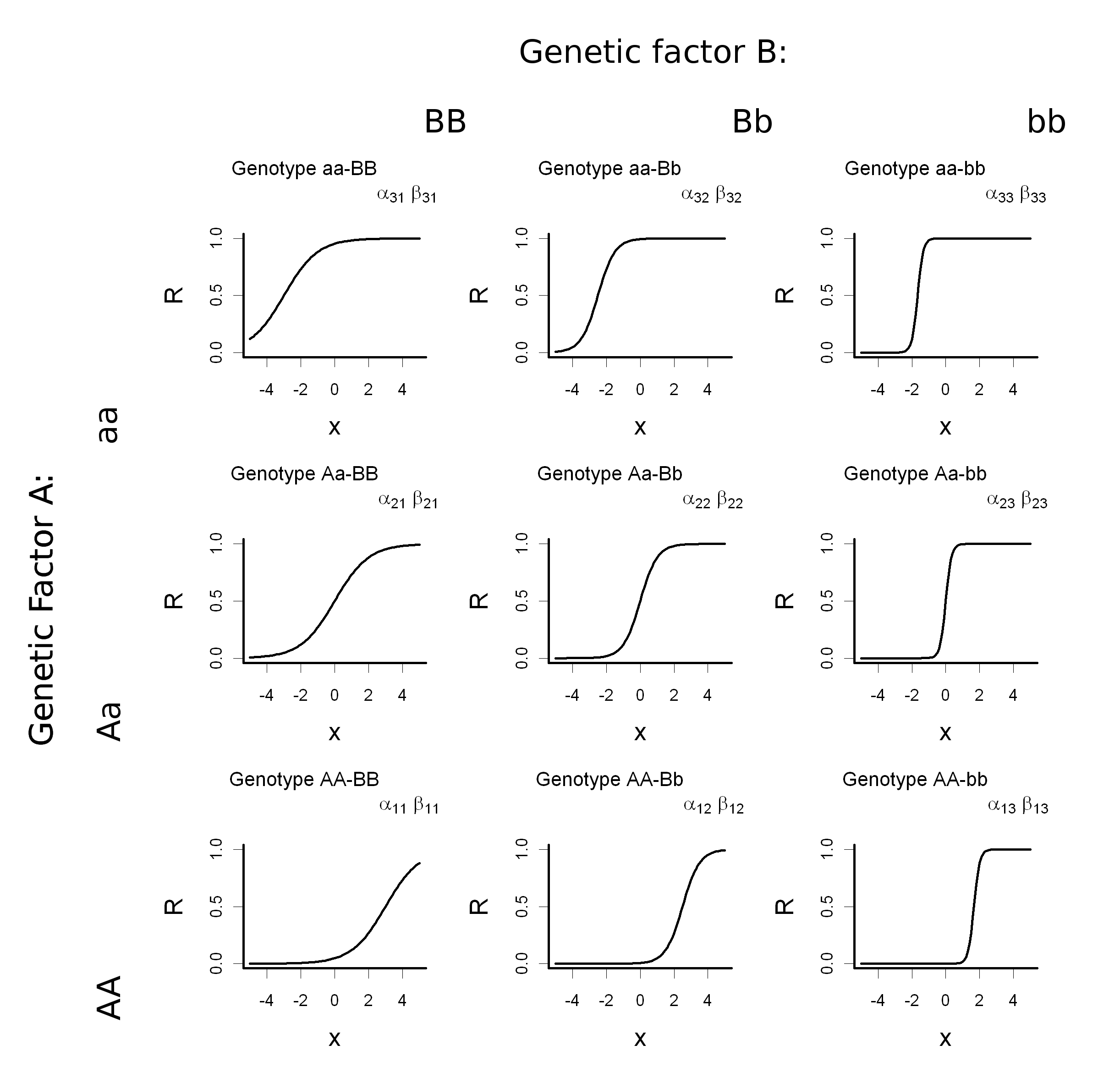} 
\caption{\textbf{Multi-logistic model applied to a two genetic-one environmental factors condition.} On the y-axis is reported the disease risk ($R$) and on the x-axis is reported the level of exposure of the environmental factor. The relationship is modelled by the Eq. \ref{eq:1}. For each combination of genetic factors there are different $\alpha_i$ and $\beta_i$ that models the relationship between environmental exposure and disease risk.}
\end{figure}

\begin{figure}[H]
\includegraphics[width=\textwidth]{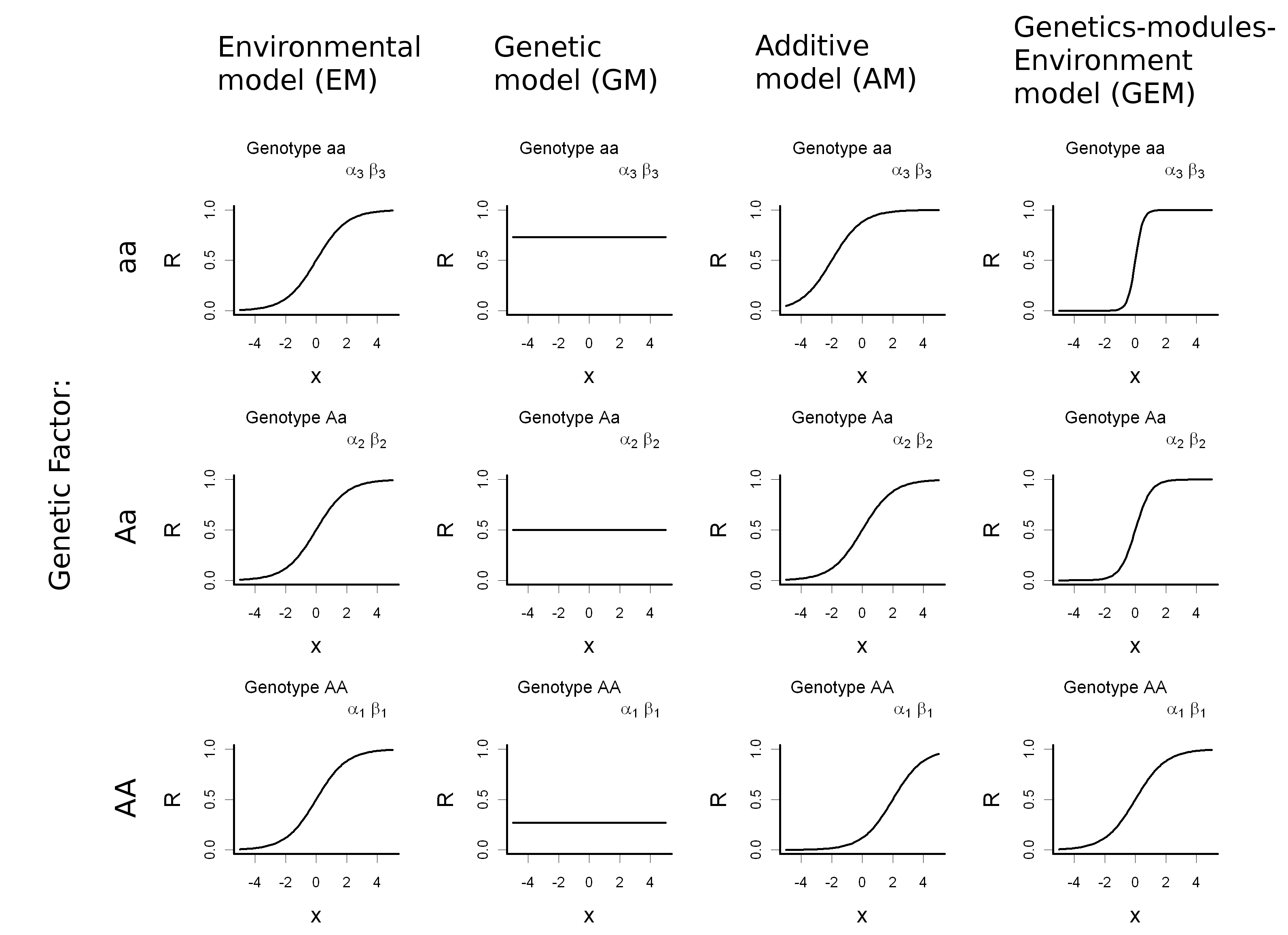} 
\caption{\textbf{Type of GxE interactions modeled by KAPS.} On the y-axis is reported the disease risk ($R$) and on the x-axis is reported the level of exposure of the environmental factor. The relationship is modelled by the Eq. \ref{eq:1}. For each combination of genetic factors there are different $\alpha_i$ and $\beta_i$ that follows the specific constraints (Table 1). In the Environmental Model (EM), the disease risk is dependent only by the environmental exposure level, thus the environment-risk relationship is the same across genotype (same slope and no shift). In the Genetic Model (GM), the disease risk depends by genetic factor only, thus the environment has no role on the disease risk (the curve is flat) while the risk is different across genotypes (height of the curve). In the third model (AM), the disease risk depends by both genetic and environmental factors; the relationship between environmental exposure and disease risk is the same in each genotype (same slope), but in each genotype there is a different basal risk (shift). In the fourth model (GEM), the genetic factor influences the relationship between environmental exposure and disease risk (slope). However, there is no different basal genetic risk (no shift).}
\end{figure}

\begin{figure}[H]
\includegraphics[width=\textwidth]{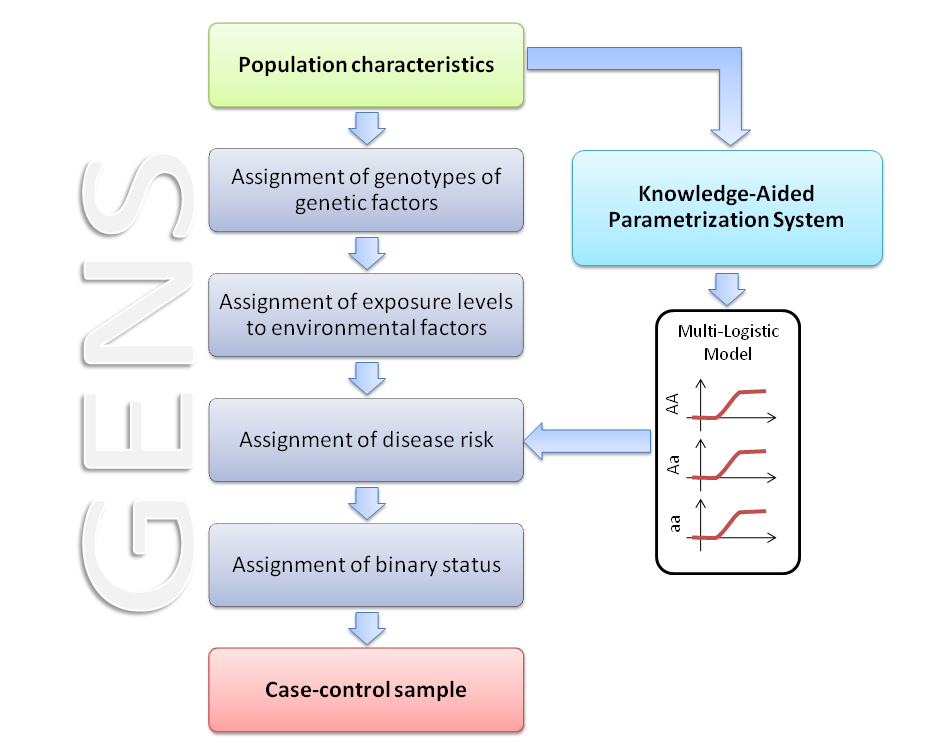} 
\caption{\textbf{Flowchart of GENS} Starting from desired population characteristics, GENS assigns to each individual the genotypes of genetic factors and the exposure levels of environmental factors. Beside, the module KAPS uses population characteristics to compute coefficients of the Multi-Logistic Model. Thus, on the basis of individual characteristics and Multi-Logistic Model, the individual disease risk is computed. The last step is the assignment of disease status to individuals (affected/not affected) according to their disease risks.}
\end{figure}

\section{Tables}

\begin{table}[H]
\begin{tabular}{|l|l|}
        \hline Interaction model & Constraints \\ \hline \hline
        Genetic Model & $\alpha_1 \leq \alpha_2 \leq \alpha_3$ and
$\beta_1=\beta_2=\beta_3=0$ \\ \hline
        Environmental Model & $\alpha_1 = \alpha_2 =
\alpha_3$ and $\beta_1=\beta_2=\beta_3=\beta \neq 0$ \\ \hline
        Gene Environment interaction Model & $\alpha_1 =
\alpha_2 = \alpha_3$ and $\beta_1 \leq \beta_2 \leq \beta_3$ \\ \hline
        Additive Model & $\alpha_1 \leq \alpha_2 \leq
\alpha_3$ and $\beta_1=\beta_2=\beta_3=\beta\neq 0$ \\ \hline
      \end{tabular}
\caption{\textbf{Relationships among the coefficients of the Multi-Logistic Model and the type of interaction.} Type of gene-environment interactions are expressed as constraints among coefficients of the Multi-Logistic Model. This approach allows to specify the type of interaction to simulate in a simple manner. Another interesting consequence is that for each type of interaction only a subset of coefficients needs to be specified.}
\end{table}

\begin{table}[H]
      \begin{tabular}{|c|p{0.75\textwidth}|}
        \hline Parameter & Description \\ \hline \hline
        $N$ & Number of individuals \\ \hline
        $G$ & Number of genetic factors \\ \hline
        $E$ & Number of environmental factors \\ \hline
        $P^G(g^a_{i_a})$ & Frequency of genotype $i_a$ of genetic factor $g^a$ ($a=1,...,G$) \\ \hline
        $P^E(x^b_{j_b})$ & Exposure probabilities, where $b=1,...,E$ and $j_b$ is an index which runs over the possible discretized values of the variable $b$ \\ \hline
        $m$ & Overall disease frequency in the population \\ \hline
        TypeOfGxe & Type of GxE: Genetic (GM), Environmental (EM), Gene Environment interaction (GEM, Additive (AM) \\ \hline
        $RR_{31}$ & Relative risk of high-risk homozygote \\ \hline
        $W$ & Model of inheritance: recessive ($W=0$), dominant ($W=1$), co-dominant ($0 < W < 1$) \\ \hline
        $\beta$ & Odds ratio of the disease risk of an individual exposed to $x_j$ with respect to one exposed to $x_j+1$ \\ \hline
      \end{tabular}
\caption{\textbf{Parameters required by GENS.} Description of parameters required by GENS in order to produce a simulated case-control sample. These parameters are translated into coefficients for the Multi-Logistic Model by the Knowledge-Aided Parametrization System.}
\end{table}

\end{document}